\documentclass[aps,prc,twocolumn,nofootinbib,amsmath,amssymb]{revtex4-2}
\usepackage[utf8]{inputenc}
\usepackage{amsmath}
\usepackage{graphicx}
\usepackage{float}
\usepackage{hyperref}
\usepackage{lipsum}
\usepackage{listings}
\usepackage{listingsutf8}
\usepackage{xcolor}
\usepackage{amssymb}
\usepackage{textcomp}
\usepackage{units}
\usepackage{bbm}
\usepackage{enumitem}  
\usepackage{url}
\usepackage[english]{isodate}
\usepackage{bm}
\usepackage{mdframed}
\usepackage[capitalise]{cleveref}
\usepackage{enumerate}
\usepackage{dsfont}
\usepackage{dcolumn}
\usepackage{physics}
\usepackage{color}
\usepackage{placeins}
\usepackage{slashed}
\usepackage{amssymb}
\usepackage{hyperref}
\usepackage[capitalise]{cleveref}
\usepackage[caption=false]{subfig}
\usepackage{braket}

\renewcommand{\Im}{\operatorname{Im}}
\renewcommand{\Tr}{\operatorname{Tr}}
\newcommand{\VI}{V_\mathrm{I}}
\newcommand{\VII}{V_\mathrm{II}}

\renewcommand{\l}{\ell}

\newcommand{\np}{$np$}

\newcommand*{\thetacm}{\theta_\mathrm{cm}}
\newcommand*{\tS}{^3S_1 \mathrm{-}^3D_1}
\newcommand*{\tP}{^3P_2 \mathrm{-}^3F_2}

\newcommand*{\LO}{LO}
\newcommand*{\NLO}{NLO}
\newcommand*{\NNLO}{N$^2$LO}
\newcommand*{\NNNLO}{N$^3$LO}

\newcommand*{\VLO}{V^{(0)}}
\newcommand*{\VNLO}{V^{(1)}}
\newcommand*{\VNNLO}{V^{(2)}}
\newcommand*{\VNNNLO}{V^{(3)}}

\newcommand*{\TLO}{T^{(0)}}
\newcommand*{\TNLO}{T^{(1)}}
\newcommand*{\TNNLO}{T^{(2)}}
\newcommand*{\TNNNLO}{T^{(3)}}

\newcommand*{\MN}{m_N}
\newcommand*{\SGT}{\sigma_\mathrm{tot}}
\newcommand*{\Tl}{T_\mathrm{lab}}
\newcommand*{\CM}{c.m.\@}

\newcommand*{\Mhi}{\Lambda_b}
\newcommand*{\pon}{p_0}

\begin{document}

\title{Perturbative computations of neutron-proton scattering observables using renormalization-group invariant $\chi$EFT up to N$^3$LO}
\author{Oliver Thim} \email{oliver.thim@chalmers.se}
\author{Andreas Ekström}
\author{Christian Forssén}
\affiliation{Department of Physics, Chalmers University of Technology, SE-412 96, Göteborg, Sweden}

\date{\today}
\noindent
\begin{abstract} 
We predict neutron-proton scattering cross-sections and polarization observables up to next-to-next-to-next-to leading order in a renormalization-group invariant description of the strong nucleon-nucleon interaction. Low-energy constants are calibrated to phase shifts, sub-leading corrections are computed in distorted-wave perturbation theory, and we employ momentum-cutoff values 500 and 2500 MeV. We find a steady order-by-order convergence and realistic descriptions of scattering observables up to a laboratory scattering energy of approximately 100~MeV. We also compare perturbative and non-perturbative calculations for phase shifts and cross sections and quantify how unitarity is gradually restored at higher orders. The perturbative approach offers an important diagnostic tool for any power counting and our results suggest that the breakdown scale in chiral effective field theory might be significantly lower than  estimates obtained in non-perturbative calculations. 

\end{abstract}

\maketitle

\newpage

\newpage
\section{Introduction}
Nuclear potentials used in \textit{ab initio}~\cite{Ekstrom:2022yea} computations of atomic nuclei~\cite{Hergert:2020bxy} are almost exclusively derived using chiral effective field theory ($\chi$EFT) \cite{Epelbaum:2008ga,Machleidt:2011zz,Hammer:2019poc} based on Weinberg power counting (WPC)~\cite{Weinberg:1990rz,Weinberg:1991um}. Such potentials~\cite{Entem:2003ft,Ekstrom:2013kea,Gezerlis:2013ipa,Piarulli:2014bda,Carlsson:2015vda,Ekstrom:2015rta,Jiang:2020the}, now derived up to the fifth chiral order~\cite{Reinert:2017usi,Entem:2017gor,Epelbaum:2019kcf}, have furnished a wide range of structure and reaction predictions across the nuclear chart~\cite{Tews:2020hgp,Tews:2022yfb}, but at the same time they grapple with the renormalization challenge inherent to chiral nuclear forces~\cite{vanKolck:2020llt}. Indeed, numerical studies~\cite{Nogga:2005hy} of the nucleon-nucleon scattering amplitude have shown that the contact operators, accounting for unresolved short-range physics, already at leading order (LO) in WPC are not sufficient to renormalize the singular nature~\cite{Frank:1971xx} of the one pion-exchange potential. Consequently, LO predictions based on WPC exhibit an unphysical dependence on the cutoff $\Lambda$ that regularizes the amount of high-momentum (or short-range) physics that is resolved. 

Several PCs leading to renormalization-group (RG) invariant nucleon-nucleon amplitudes have been proposed in the past two decades~\cite{PavonValderrama:2005uj,PavonValderrama:2005gu,PavonValderrama:2011fcz,Long:2013cya,PavonValderrama:2016lqn,Valderrama:2009ei,Birse:2005um,Long:2012ve,SanchezSanchez:2017tws,PhysRevC.85.034002,Yang:2016brl,Mishra:2021luw,Peng:2021pvo}. They can collectively be referred to as modified Weinberg power countings (MWPCs).
However, we typically know very little about their predictive power for nuclei beyond the lightest-mass systems~\cite{Song:2016ale}. The one exception is the recent study by \citet{Yang:2020pgi} that presented the first \textit{ab initio} predictions of binding energies in $^4$He, $^6$Li, and $^{16}$O using $\chi$EFT potentials up to next-to-leading order (NLO) in several different MWPCs. The calculations in that work revealed an $\alpha$-decay instability in the ground states in $^6$Li and $^{16}$O. Subsequent analyses brought forward probable causes for this instability as originating in ($i$) overfitting of the low-energy constants (LECs) that parameterize the short-range interactions~\cite{Thim:2023fnl} and ($ii$) underestimating the importance of few-nucleon forces~\cite{Yang:2021vxa} at LO in MWPC. 

The notable absence of MWPC-based predictions for heavier-mass nuclei is likely due to a variety of factors. Firstly, potentials based on WPC are easier to implement in present \textit{ab initio} computer codes as one straightforwardly sum leading and sub-leading corrections to the potential before solving the Schrödinger equation, whereas in MWPC sub-leading corrections should be added in perturbation theory~\cite{Long:2007vp}. Secondly, there exists several widely available computer codes for evaluating matrix elements of chiral nucleon-nucleon and three-nucleon potentials, as well as currents, to very high orders in WPC. Finally, it is currently prohibitively costly to converge \textit{ab initio} predictions of nuclear properties at the large values of the cutoff required for analyzing RG-invariance in MWPC.

In light of these facts we certainly see the utility of WPC, which might provide a consistent EFT framework provided that renormalization is interpreted in a fashion where the cutoff never exceeds the order of the breakdown scale ~\cite{Epelbaum:2006pt,Epelbaum:2009sd,Epelbaum:2018zli}. However, the existence of MWPCs, where renormalization does allow for the cutoff to be taken far beyond the breakdown scale, calls for a continued effort. We note that it was shown in Ref.~\cite{Gasparyan:2022isg} that one can encounter so-called exceptional points in the cutoff domain beyond the breakdown scale for which the cutoff independence of the phase shifts significantly deteriorates. Given the fundamental importance of RG-invariance it should be seriously explored whether MWPC approaches can furnish a realistic and predictive framework for \textit{ab initio} nuclear physics.

In this paper, we contribute to the meager list of quantitative predictions grounded in RG-invariant formulations of $\chi$EFT. To the best of our knowledge, and somewhat surprisingly, nucleon-nucleon scattering observables have not been computed in MWPC beyond LO~\cite{Epelbaum:2006pt}. Here, we present predictions for integrated and differential cross-sections, as well as polarization observables, for elastic neutron-proton (\np) scattering up to next-to-next-to-next-to-leading order (\NNNLO) in the MWPC of Long and Yang~\cite{Long:2012ve,PhysRevC.84.057001,PhysRevC.85.034002}, where higher-order corrections to the potential are treated perturbatively~\cite{Nogga:2005hy,Long:2007vp}. This work serves as an important step in the development and uncertainty quantification of any model of the nuclear interaction~\cite{Furnstahl:2015rha,Wesolowski_2019,Wesolowski:2021cni,Svensson:2022,Svensson:2023twt}. 

In \cref{sec:formalism} we review how to construct potentials in the PC of Long and Yang, describe how to numerically compute the scattering amplitude in distorted-wave perturbation theory, and explain how we calibrated LEC values. In \cref{sec:observables} we present results for scattering observables up to \NNNLO, and we summarize and conclude in \cref{sec:conclusions}.

\section{Formalism\label{sec:formalism}}

In $\chi$EFT, scattering amplitudes are expanded in a dimensionless ratio $(Q/\Mhi)^{\nu}$. Here, $\nu$ indicates the chiral order, $\Mhi$ is the underlying high-momentum scale of $\chi$EFT, and $Q$ denotes the relevant low-energy scale. For nucleon-nucleon scattering, we assume $Q\approx \text{max}(p,m_\pi)$, where $p$ is the relative momentum in the center of mass (c.m.) frame of the interacting nucleons, and the pion mass $m_\pi$ is the relevant low-energy mass scale. In this work we adopt a nomenclature where \LO{} scales as $\left(Q/\Mhi\right)^0$ while sub-leading orders are denoted by their relative scaling to LO. As such, \NLO{}  scales as $\left(Q/\Mhi\right)^1$, next-to-next-to-leading order (\NNLO) as $\left(Q/\Mhi\right)^2$ and so on. In what follows, we summarize relevant details regarding the MWPC that we use in this work, define the potential terms $V^{(\nu)}$ entering at each chiral order, and explain how we performed the perturbative calculations of scattering amplitudes.

\subsection{The nucleon-nucleon interaction potential in the Long and Yang power counting \label{sec:PC_and_pot}}
We employ the MWPC of Long and Yang~\cite{Long:2012ve,PhysRevC.85.034002,Long:2011qx,Long:2007vp}, which adheres to the following overarching principles:
\begin{itemize}
    \item The chiral order of a pion-exchange diagram, along with the necessary counterterms for renormalizing pion loops, is determined by the naive dimensional analysis (NDA) of its non-analytic part. This follows the same principle as in Weinberg Power Counting (WPC).
    \item Counterterms are promoted to lower chiral order only when needed to fulfill the requirement of RG-invariance.
    \item All corrections to the potential beyond LO are included perturbatively to obtain RG-invariant amplitudes.
\end{itemize}

One-pion exchange (OPE) enters at LO in $\chi$EFT and must be treated non-perturbatively, at least  in the low partial waves where it is sufficiently strong. The singular nature of OPE is increasingly alleviated by the centrifugal barrier. Thus, at some point in the partial-wave expansion there is sufficient angular momentum $\ell$ to furnish a perturbative treatment of OPE~\cite{Birse:2005um, PhysRevC.99.024003, Peng:2020nyz} and consider it sub-leading. 

At \LO{} in the MWPC by Long and Yang, the OPE potential $V^{(0)}_{1\pi}$ is considered non-perturbative in the $^1S_0$, $^3P_0$, $^1P_1$, $^3P_1$, $\tS$ and $\tP$ channels. OPE is attractive in $^3P_0$ and $^3P_2$. Renormalization requires promotion of counterterms to the corresponding channels of the \LO{} contact potential $V^{(0)}_{\mathrm{ct}}$~\cite{Nogga:2005hy}, thereby extending it beyond the canonical non-derivative $^1S_0$ and $^3S_1$ counterterms. At sub-leading orders ($\nu>0$), two pion-exchange, $V_{2\pi}^{(\nu)}$, as well as higher-order contact potentials, $V_\text{ct}^{(\nu)}$, enter perturbatively according to the principles presented in the beginning of this subsection. The contributions to the potential up to \NNNLO{} in the $^1S_0$, $^3P_0$, $^1P_1$, $^3P_1$, $\tS$ and $\tP$ channels are listed in the third column of \cref{tab:potentials_PC} labeled ''non-perturbative (at LO) channels''.

See~\cref{app:LY_and_pot} for detailed expressions of the potentials appearing in \cref{tab:potentials_PC}. Following Long and Yang, we do not consider any higher-order corrections to OPE and employ potential expressions where pion loops are treated in dimensional regularization. For the sub-leading two-pion exchange potential $V^{(3)}_{2\pi}$ we use pion-nucleon LECs $c_1,c_3,c_4$ with central values from the Roy-Steiner analysis in Ref.~\cite{Siemens:2016jwj}. 

\begin{table}
    \caption{Potential contributions at each in  channels where OPE is treated non-perturbatively (column three) and perturbatively (column four). Detailed expressions for the potentials can be found in \cref{app:LY_and_pot}.}
\centering
\begingroup
\renewcommand{\arraystretch}{1.35} 
\begin{tabular}{l|c|c|c}
    & & non-perturbative (at LO) & purely perturbative \\
    order & potential & channels & channels \\
    \toprule
    LO & $V^{(0)}$ & $V^{(0)}_{1\pi} + V^{(0)}_{\mathrm{ct}}$ & 0 \\
    NLO & $V^{(1)}$ & $V^{(1)}_{\mathrm{ct}}$ & $V^{(0)}_{1\pi}$ \\
    \NNLO{} & $V^{(2)}$ & $V^{(2)}_{2\pi} + V^{(2)}_{\mathrm{ct}}$ & 0 \\
    \NNNLO{} & $V^{(3)}$ & $V^{(3)}_{2\pi} + V^{(3)}_{\mathrm{ct}}$ & $V^{(2)}_{2\pi}$ \\ 
    \hline
\end{tabular}
\endgroup
\label{tab:potentials_PC}
\end{table}

Let us now turn to the channels with $\l>1$ (and without any coupling to $\l \leq 1$). For these channels we consider OPE to be perturbative and consequently set it to zero at \LO. We follow Ref.~\cite{PhysRevC.99.024003} and suppress two-pion exchanges by the same chiral power as OPE. Up to \NNNLO, there are no contact potentials in the perturbative channels, and the contributions are listed in the last column of \cref{tab:potentials_PC}.
Other suggestions for the PC in perturbative channels are discussed by, e.g.,~\citet{PavonValderrama:2016lqn}. 

\subsection{A perturbative treatment of nucleon-nucleon scattering amplitudes\label{sec:dwpt}}
The perturbative computation of nucleon-nucleon scattering amplitudes proceeds in two steps. First, we solve the Lippmann-Schwinger (LS) equation for the \LO{} amplitude in the $^1S_0$, $^3P_0$, $^1P_1$, $^3P_1$, $\tS$ and $\tP$ channels. Note that the \LO{} potential is identically zero in all other channels. Second, we perturbatively include higher-order potential corrections to the amplitude, accounting for the distortion due to the non-perturbative \LO{} solution where necessary. In the following, we explain this procedure in detail, see also Refs.~\cite{Peng:2020nyz,PhysRevC.85.034002,Long:2012ve}.

The neutron-proton Hamiltonian in the center-of-mass (\CM) frame can be written
\begin{equation}
    H = \frac{\bm{p}^2}{m_N} + V_\mathrm{I} + V_\mathrm{II},
    \label{eq:H_I_II}
\end{equation}
where $\bm{p}$ denotes the \CM\ momentum and $m_N = 2 m_n m_p/(m_n+m_p)$ the nucleon mass. The projectile energy in the laboratory frame will be denoted $\Tl$. Furthermore, $V_\mathrm{I}$ denotes the \LO{} potential, and $V_\mathrm{II}$ denotes the sum of all sub-leading potentials, which formally can be infinitely many. The PC helps us identify important and less important contributions to the scattering amplitude $T$ and therefore facilitates a meaningful truncation of $V_\mathrm{II}$. With the notation for the chiral potentials $V^{(\nu)}$ introduced in \cref{sec:PC_and_pot}, $\VI$ and $\VII$ read
\begin{align}
    \VI  &= \VLO, \\
    \VII &= \sum_{\nu=1}^\infty V^{(\nu)}. 
    \label{eq:VII}
\end{align}

The \LO{} amplitude, $\TLO$, is obtained (non-perturbatively) by solving the LS-equation
\begin{equation}
    \TLO = \VLO + \VLO G^+_0\TLO,
    \label{eq:LS_op}
\end{equation}
where the free resolvent is given by
\begin{equation}
    G^+_0 = \left(E -H_0 + i\epsilon\right)^{-1},
\end{equation}
and $H_0 = \bm{p}^2/m_N$. We use a notation where we suppress the explicit dependence on the \CM\ scattering energy, $E$, for the resolvents and amplitudes.

In WPC, higher-order corrections are accounted for non-perturbatively by solving the LS-equation for the sum $\VI + \VII$. In MWPC, however, potentials beyond \LO{}, i.e., the corrections ($\VII$), enter in perturbation theory to obtain RG invariant results~\cite{Long:2007vp}. Indeed, higher-order corrections should be amenable to a perturbative treatment. If not, they are non-perturbative in nature and belongs at \LO. 

\begin{figure*}[ht]
    \centering
    \includegraphics[width=0.65\textwidth]{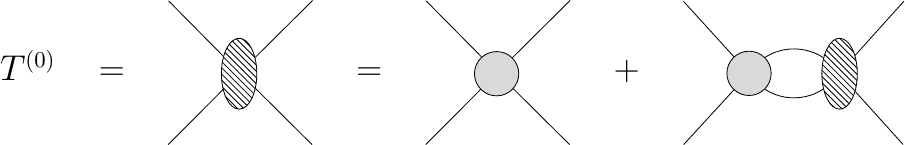}
    \includegraphics[width=0.94\textwidth]{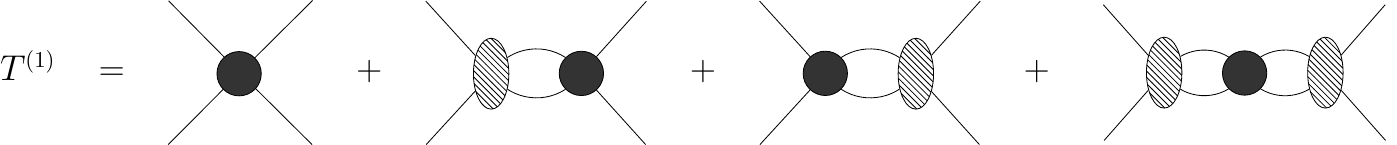}
    \caption{Diagrammatic representation of the LO neutron-proton amplitude $T^{(0)}$ (hatched oval), obtained by solving the LS-equation, as well as the first correction $T^{(1)}$ given in \cref{eq:TNLO}. The grey (black) solid blobs represent the potentials $V^{(0)}$ ($V^{(1)}$).}
    \label{fig:diagrams}
\end{figure*}

Distorted-wave perturbation theory has been applied to compute scattering amplitudes in several previous studies, see, e.g., Refs.~\cite{Valderrama:2009ei,Peng:2020nyz,Long:2012ve,PhysRevC.85.034002,Long:2011qx,Barford:2002je}. The perturbation series for the scattering amplitude can be derived and expressed in various ways. The one that we find most instructive follows Refs.~\cite{Newton:1982qc,Hussein_Canto_2012}. First, using the two-potential trick, the $T$-operator for the Hamiltonian in \cref{eq:H_I_II} is written in the form
\begin{equation}
    T  =  \TLO +  \Omega^\dagger_- V_\mathrm{II} \sum_{n=0}^\infty \left(G^+_1V_\mathrm{II}\right)^n \Omega_+,
    \label{eq:full_DWB_moller}
\end{equation}
where the M\o ller wave operators are defined as
\begin{align}
    \Omega_+ &= \mathds{1} + G^+_0\TLO,     \label{eq:moller1} \\
    \Omega^\dagger_- &= \mathds{1} + \TLO G^+_0,
    \label{eq:moller2}
\end{align}
and the full \LO{} resolvent reads
\begin{equation}
    G^+_1 = \Omega_+ G^+_0.
    \label{eq:G1}
\end{equation}
Inserting \cref{eq:VII} in \cref{eq:full_DWB_moller} gives for the full $T$-operator
\begin{equation}
  T = \TLO +  \Omega^\dagger_- \left[\sum_{\nu=1}^\infty V^{(\nu)}\right] \sum_{n=0}^\infty \left[ G^+_1\left(\sum_{\nu'=1}^\infty V^{(\nu')}\right)\right]^n \Omega_+.
\end{equation}
Expanding both sums and organizing terms according to their chiral orders $\nu$ yields the expressions for the first-, second-, and third-order \textit{corrections} to the \LO{} amplitude as
\begin{align}
    \TNLO &= \Omega^\dagger_-\VNLO\Omega_+ \label{eq:TNLO}\\
    \TNNLO &= \Omega^\dagger_-\left(\VNNLO  +   \VNLO G^+_1 \VNLO\right) \Omega_+\label{eq:TN2LO}\\
    \TNNNLO &= \Omega^\dagger_-\Big(\VNNNLO +   \VNNLO G^+_1\VNLO +  \VNLO G^+_1\VNNLO + \nonumber\\ &+ \VNLO G^+_1\VNLO G^+_1\VNLO \Big)\Omega_+.\label{eq:TN3LO}
\end{align}
A diagrammatic representation of amplitudes up to \NLO{} is presented in \cref{fig:diagrams}.
Note that the full amplitude at, e.g., third order (\NNNLO) is given by the sum $T^{(0)}+T^{(1)}+T^{(2)}+T^{(3)}$. Clearly, the distorted-wave corrections in \cref{eq:TNLO,eq:TN2LO,eq:TN3LO} simplify dramatically when applied to the channels where OPE is perturbative such that $\TLO = 0$, $\Omega_+ = \mathds{1}$, and $\Omega^\dagger_- = \mathds{1}$. In these channels we therefore recover ordinary perturbation theory.

The distorted-wave corrections to the amplitudes $T^{(\nu>0)}$ can alternatively be obtained as solutions to a set of modified LS-type equations, discussed in more detail in Refs.~\cite{Griesshammer:2021zzz,Vanasse:2013sda}, which read
\begin{equation}
    T^{(\nu)} = V^{(\nu)} + \sum_{i=1}^\nu V^{(i)}G^+_0 T^{(\nu-i)} + V^{(0)} G^+_0 T^{(\nu)}.
    \label{eq:modified_LS}
\end{equation}
We use this formulation to verify our numerical implementation of \cref{eq:TNLO,eq:TN2LO,eq:TN3LO}. We note that the alternative approach of modified LS-equations requires a matrix inversion at each order, whereas the distorted-wave approach requires matrix multiplications only. However, the number of matrix multiplications increases rapidly as the chiral order is increased. For example, at $\nu=10$, \cref{eq:TNLO,eq:TN2LO,eq:TN3LO} require an order of magnitude more matrix multiplications than the modified LS equations in~\cref{eq:modified_LS}. In this study we only go to $\nu=3$ for which the number of matrix multiplications of the two formulations are similar. 

\subsection{Numerical implementation}
We project potentials and amplitudes to a partial-wave basis of states $\ket{p,\l,s,j}$ following the prescription in Ref. \cite{Erkelenz:1971caz}\footnote{Note the mistake in Eq. (4.22) pointed out in Ref. \cite{Machleidt:2011zz}.}. Here, $p = |\bm{p}|$, while $s,\l,j$ denote the quantum numbers of the two-nucleon spin, orbital angular momentum, and total angular momentum, respectively. Partial-wave matrix elements are denoted by
\begin{equation}
     V^{js}_{\l'\l}(p',p) = \braket{p',\l',s,j|V|p,\l,s,j},
\end{equation}
where the conserved quantum numbers $s$ and $j$ are given as superscripts.

In the LS-equation, as well as in \cref{eq:TNLO,eq:TN2LO,eq:TN3LO}, infinite momentum integrals appear and all potentials are regulated according to
\begin{equation}
    V^{js}_{\l'\l}(p',p) \to f_{\Lambda}(p') \ V^{js}_{\l'\l}(p',p) f_{\Lambda}(p),
\label{eq:inc_reg}
\end{equation}
where we choose a regulator function
\begin{equation}
    f_{\Lambda}(p) = \exp\left[-\frac{p^6}{\Lambda^6}\right]
\end{equation}
at all orders up to \NNNLO. In the calibration of the LECs, we use the cutoff values $\Lambda = 500$~MeV and $\Lambda = 2500$~MeV.

Using~\cref{eq:moller1,eq:moller2,eq:G1}, the terms in \cref{eq:TNLO,eq:TN2LO,eq:TN3LO} can be expanded to sums of products of the form $A_1G^+_0A_2$, of varying length. The $A_i$'s are either $T^{(0)}$ or $V^{(\nu)}$ with $\nu = 1,2,3$. For example, the \NLO{} correction in \cref{eq:TNLO} reads
\begin{align}
    T^{(1)} &= V^{(1)} + T^{(0)}G^+_0V^{(1)} + V^{(1)}G^+_0T^{(0)} \nonumber \\ &+T^{(0)}G^+_0V^{(1)}G^+_0T^{(0)}.
\end{align}
Clearly, the fundamental matrix elements that need to be evaluated at sub-leading orders are always of the form\begin{equation}
    \bra{p',\l'}A_1 G^+_0 A_2 \ket{p,\l},
    \label{eq:fund_G0_prod}
\end{equation}
where we omit the $s$ and $j$ quantum numbers that are identical for the ket and the bra. In \cref{app:numerical} we show how to evaluate \cref{eq:fund_G0_prod} using ordinary matrix products and Gauss-Legendre quadrature. Longer products, e.g., of the form $A_1G^+_0A_2G^+_0A_3$, are straightforwardly reduced to the form in \cref{eq:fund_G0_prod} by the associativity of matrix products. Knowing this, and the distributive property with respect to addition, we can also reduce the computational complexity of evaluating the perturbation series for $T$ by computing and storing the composite operators $\Omega^\dagger_-$, $\Omega_+$, and $G^+_1$. 

For separable potentials of Yamaguchi type \cite{Yamaguchi:1954mp}, both the distorted-wave series and the LS equation can be solved analytically. We exploit this to verify our numerical implementation and to inspect the stability of the perturbative expansion. Numerical and analytical results for semi-realistic and separable Yamaguchi potentials in the $^1S_0$ and $\tS$ channels agree to at least single precision.

\subsection{Calibrating the low-energy constants\label{sec:lecs}}
Our focus in this work is to predict and analyze the description of \np{} scattering observables in MWPC and specifically the PC of Long and Yang. To enable quantitative calculations, we calibrate the values of the unknown LECs using the same approach as Long and Yang, i.e., by tuning the contact LECs to achieve a good reproduction of the low-energy Nijmegen phase shifts~\cite{Stoks:1993tb} at selected scattering energies. 

\begin{figure*}
    \centering
    \includegraphics[width=0.94\textwidth]{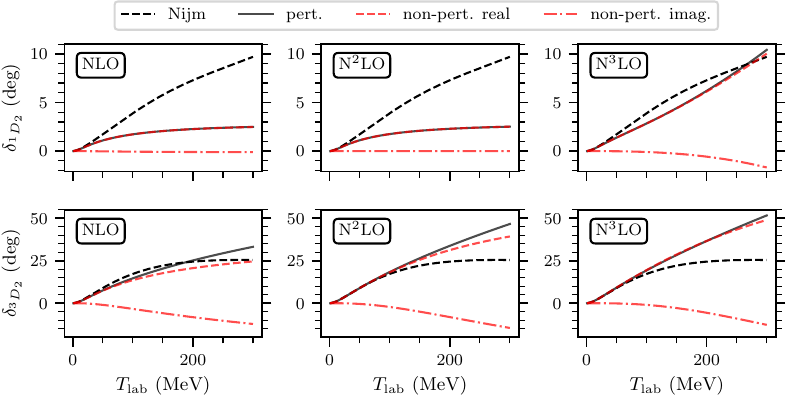}
    \caption{\np{} scattering phase shifts in the $^1D_2$ (top row) and $^3D_2$ (bottom row) channels at \NLO, \NNLO, and \NNNLO{} using a momentum cutoff $\Lambda=2500$ MeV. Phase shifts computed using the perturbative method are shown with black solid lines. The red dashed and dot-dashed lines show the real and imaginary parts, respectively, of the phase shift computed by summing the $T$-matrix contribution and using the non-perturbative relation between phase shifts and the $S$-matrix. The black dashed lines show phase shifts from the Nijmegen analysis \cite{Stoks:1993tb}.}
    \label{fig:phase_pert_nonpert}
\end{figure*}

Before discussing the details of the calibration, it is important to remember that the order-by-order amplitudes
\begin{equation}
T = \TLO+\TNLO+\TNNLO +\ldots
\end{equation}
are computed perturbatively and their sum is unitary only up to perturbative corrections. 
We therefore consider it natural to compute phase shifts perturbatively as well and proceed in this work by expanding the \np{} $S$-matrix and match to chiral orders, see~\cref{app:pert_phase} for details. In doing so the phase shifts are real by construction. If one instead solves for the partial-wave $S$-matrix non-perturbatively from the order-by-order sum of $T^{(\nu)}$ amplitudes, the corresponding phase shifts will have a non-zero imaginary part that increases with scattering energy. Indeed, \cref{fig:phase_pert_nonpert} shows phase shifts computed perturbatively and non-perturbatively in the two channels $^1D_2$ and $^3D_2$. Note that there are no LECs that need to be calibrated in these channels at the orders considered in this work. The imaginary part of the non-perturbative phase shift increases with scattering energy. As that happens, the real part of the phase shift and the (real-valued) perturbative phase shift differ progressively. This is consistent with observations in Ref.~\cite{Odell:2023gfd} and the perturbative and non-perturbative approaches of defining the phase shifts are consistent within the theoretical uncertainty due to omitted and higher-order amplitudes. An alternative approach would therefore be to work with the real part of the perturbative amplitude as done in Refs.~\cite{Kaiser:1997mw,Gasser:1990bv}.

In the calibration of LECs, we do not account for uncertainties stemming from the Nijmegen phase shifts or the truncation of the $\chi$EFT expansion. While we are aware of the potential risk of overfitting in doing so, we opted for a simple approach to establish a first quantitative potential and a baseline understanding. The application of Bayesian inference methods~\cite{Wesolowski_2019,Wesolowski:2021cni,Svensson:2022} to quantify the posterior probability distributions for the values of the LECs in MWPC~\cite{Thim:2023fnl}, though more robust, requires considerably more efforts. In this work, we focus on studying the effectiveness of MWPC for realistic description of elastic \np\ scattering. 

The $T_\mathrm{lab}$ values of the Nijmegen phase shifts used as calibration data are listed in \Cref{tab:LECs_fit_points} for each channel and order. The calibrated LECs up to \NNNLO{} are compiled in \cref{tab:LYvK_LECs_up_to_N3LO} in \cref{app:LY_and_pot}. We use a naming-convention where capital letters $C,D,E,\ldots$ denote LECs with dimension MeV$^{-2}$, MeV$^{-4}$, MeV$^{-6},\ldots$, respectively.

Each LEC receives perturbative corrections at subsequent orders from where it was first introduced. As an example, the \LO{} LEC $C_{^1S_0}$ is expanded into contributions
\begin{equation}
    C_{^1S_0} = C^{(0)}_{^1S_0} + C^{(1)}_{^1S_0} + C^{(2)}_{^1S_0} +\dots,
\end{equation}
where the superscript enumerates the perturbative correction and not the chiral order. In the following we will exemplify the calibration procedure by discussing in detail how we calibrated the LECs in the $^1S_0$ channel.

\begin{table}
    \caption{Laboratory scattering energies $T_\mathrm{lab}$ (in MeV) of the Nijmegen phase shifts~\cite{Stoks:1993tb} used to calibrate the values of the LECs at each chiral order. In total, we employed 33 single-energy phase shifts---the same as the total number of contact LECs in the chiral expansion of Long and Yang up to \NNNLO{}.}
\centering

\begingroup
\renewcommand{\arraystretch}{1.1} 
\begin{tabular}{c|c|c|c|c}
    Channel & \LO & \NLO & \NNLO & \NNNLO \\
    \hline\hline
    $^1S_0$ & 5 & 5, 25 & 5, 25, 50 & 5, 25, 50, 75 \\
    $^3P_0$ & 25 &-& 25, 50 & 75, 100 \\
    $^1P_1$ & -&- &50&50 \\
    $^3P_1$ & -&-& 50& 50 \\
    $\tS$ &  $^3S_1: 30$ & -& $^3S_1: 30,50.$ & $^3S_1: 30,50.$\\
    &  & & $\epsilon_1: 50$ & $\epsilon_1: 50$\\
    $\tP$ &  $^3P_2: 30$ & - & $^3P_2: 30,50.$& $^3P_2: 30,50.$\\
    &  & & $\epsilon_2: 50$& $\epsilon_2: 50$\\
    \hline
\end{tabular}
\endgroup
\label{tab:LECs_fit_points}
\end{table}

At \LO{} we calibrate the LEC $C^{(0)}_{^1S_0}$ such that the \LO{} $^1 S_0$ phase shift, $\delta^{(0)}$, reproduces the Nijmegen phase shift at $\Tl = 5$~MeV. Two LECs are present in the $^1S_0$ channel of the \NLO{} potential: $D^{(0)}_{^1S_0}$ and $C^{(1)}_{^1S_0}$. The latter is a perturbative correction to the \LO{} LEC. These two LECs are calibrated such that the \LO{} phase shift plus the perturbative \NLO{} correction, i.e., $\delta^{(0)} + \delta^{(1)}$, reproduce the Nijmegen phase shifts at $\Tl = 5$ and 25~MeV. The role of $C^{(1)}_{^1S_0}$ is to ensure that the \NLO{} correction vanishes for $\Tl = 5$~MeV. 
At \NNLO{} we have the LECs $C^{(2)}_{^1S_0},\ D^{(1)}_{^1S_0}, \ E^{(0)}_{^1S_0}$ calibrated to phase shifts at energies $\Tl = 5,25$ and 50~MeV. Finally, at \NNNLO{} the LECs $C^{(3)}_{^1S_0},\ D^{(2)}_{^1S_0}, \ E^{(1)}_{^1S_0}, \ F^{(0)}_{^1S_0}$ are calibrated to reproduce the phase shifts at $\Tl = 5,25,50$ and 75~MeV. An analogous scheme is employed for the remaining partial waves and LECs. We calibrate all LECs for two different momentum cutoffs: $\Lambda = 500$ and 2500~MeV. 

\begin{figure*}
    \centering
    \includegraphics[width=0.94\textwidth]{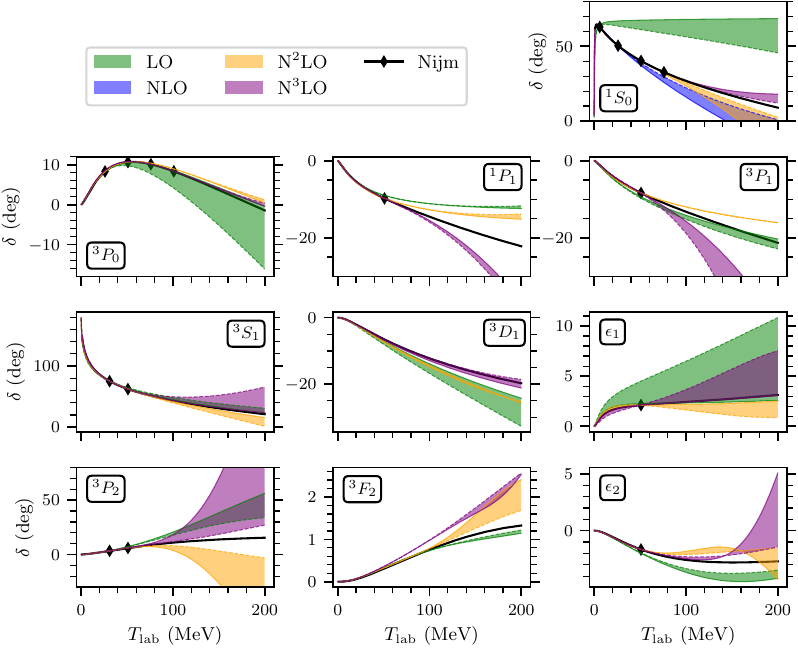}
    \caption{Phase shifts in the channels where OPE is non-perturbative and the amplitudes are computed using full distorted-wave perturbation theory. The bands indicate the envelope of the variation due to the two different cutoff values; 500 MeV (dashed line) and 2500 MeV (solid line). Note that \LO{} and \NLO{} results coincide for all channels except $^1S_0$, which is why the blue \NLO{} band appears to be missing in several panels. The black solid lines show phase shifts from the Nijmegen analysis \cite{Stoks:1993tb} and the diamond markers indicate the calibration data at $T_\mathrm{lab}$ values from \cref{tab:LECs_fit_points}.}
    \label{fig:non_pert_phase_shifts}
\end{figure*}

For the channels where OPE is perturbative there are no LECs present that need to be calibrated. As a consistency check we compute and reproduce the scattering phase shifts of Ref.~\cite{PhysRevC.99.024003}. \Cref{fig:non_pert_phase_shifts} shows our fit of the phase shifts in the channels where OPE is non-perturbative. The bands indicate the variation due to the two different cutoff values. There is an overall order-by-order convergence in all channels up to around $\Tl = 100$~MeV and we can reproduce the known results of \cite{Long:2012ve,PhysRevC.85.034002,Long:2011qx}. The degree of cutoff sensitivity varies notably among different channels. For instance, channels like $^1P_1$ and $^3F_2$ show minimal sensitivity to the cutoff value, while $^3P_2$ and $\epsilon_1$ demonstrate a more pronounced dependency. The calibration in the $^3P_0$ channel was particularly challenging at the higher chiral orders and the calibration energies needed to be shifted to relatively high values at \NNNLO{}, as seen in \cref{tab:LECs_fit_points}.

\section{Neutron-Proton Scattering Observables\label{sec:observables}}

Here we predict selected \np{} scattering observables up to $T_\mathrm{lab} \approx 100$~MeV using the potentials that were defined and calibrated in~\cref{sec:formalism}. We compute scattering observables from the partial-wave amplitudes by first constructing the spin-scattering matrix, $M$,  by~\cite{Blatt:1952zz,Glockle,Newton:1982qc}
\begin{align}
    M^{s}_{m'_s m_s}&(\pon,\thetacm,\phi) = \frac{\sqrt{4\pi}}{2i\pon} \sum_{j,\l,\l'} i^{\l-\l'} (2j+1)\sqrt{2\l+1}  \nonumber \\ & \times 
     \begin{pmatrix} \l' & s & j \\ m_s-m'_s & m'_s & -m_s \end{pmatrix}  \begin{pmatrix} \l & s & j \\ 0 & m_s& -m_s\end{pmatrix} \label{eq:M_matrix} \\ \nonumber
    &\times Y^{\l'}_{m_s -m'_s}(\thetacm,\phi) \left(S^{(\nu)js}_{\l'\l}(\pon,\pon)-\delta_{\l'\l}\right).
\end{align}
The angles $\thetacm\in[0,\pi]$ and $\phi\in[0,2\pi]$ are the polar and azimuthal scattering angles, respectively where the latter is set to zero by cylindrical symmetry. The on-shell scattering momentum, $\pon$, is given from the laboratory scattering energy $T_\mathrm{lab}$ using \cref{eq:q_on_shell} in \cref{app:numerical}. We compute $S^{(\nu)js}_{\l'\l}(\pon,\pon)$, i.e., the $S-$matrix for a potential up to some chiral order $\nu$, by summing the perturbatively computed $T$-matrix amplitudes to order $\nu$. Using the conventions applied in this work, the partial-wave relation between the on-shell $S$- and $T$-matrix elements is thus given by
\begin{align}
    &S^{(\nu)js}_{\l'\l}(\pon,\pon)= \delta_{\l'\l}-i \pi m_N \pon\nonumber \\ &\times\left[T^{(0)js}_{\l'\l}(\pon, \pon) + \dots + T^{(\nu)js}_{\l'\l}(\pon, \pon)\right].
    \label{eq:S-T-nu}
\end{align}
Note that we do not need to compute phase shifts as a middle step, but rather use the amplitudes directly. This means that the perturbative phase shift computations only indirectly influence the observables through the LECs.
We focus our discussion on the differential \np\ scattering cross section and two selected polarizations, and calculate these from the spin-scattering matrix as
\begin{align}
    \frac{d\sigma}{d\Omega} &= \frac{1}{4}\Tr{M M^\dagger} 
    \label{eq:DSG}\\
    \frac{d\sigma}{d\Omega} \times P_b &= \frac{1}{4}\Tr{M\bm{\sigma}_{1n} M^\dagger} \label{eq:PB}\\
    \frac{d\sigma}{d\Omega} \times A_{yy} &= \frac{1}{4}\Tr{M\bm{\sigma}_{1n} \bm{\sigma}_{2n} M^\dagger} 
    \label{eq:AYY}
\end{align}
where $\bm{\sigma}_{in} \equiv \bm{\sigma}_i\cdot \hat{\bm{n}}$ for nucleon $i$, $\bm{\sigma}_i$ is the Pauli spin matrices, and $\hat{\bm{n}}$ is normal to the scattering plane.

\begin{figure*}
    \centering
    \includegraphics[width=0.94\textwidth]{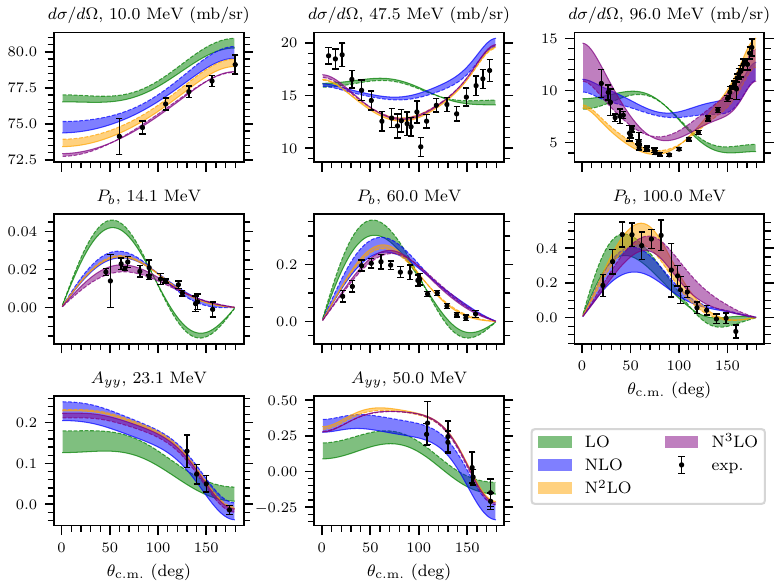}
    \caption{Selection of \np\ scattering observables in the energy interval $T_\mathrm{lab} = 10$ to 100~MeV. Experimental data from Refs. \cite{Granada_1,Granada_2}. The bands indicate cutoff variation in the same way as in \cref{fig:non_pert_phase_shifts}.}
    \label{fig:np_obs_ang}
\end{figure*}

\Cref{fig:np_obs_ang} shows our prediction for these scattering observables in the energy range $T_\mathrm{lab} = 10$ to 100~MeV for the two cutoffs $\Lambda = 500$~MeV and $\Lambda = 2500$~MeV. For the lower scattering energies ($T_\mathrm{lab} \lesssim 60$~MeV) we observe an order-by-order improvement for all considered observables. Interestingly, the \NNNLO{} predictions do not always perform better, but in general performs at least as well as \NNLO. Indeed, for $T_\text{lab}\approx$ 100 MeV (rightmost panels of \cref{fig:np_obs_ang}), it appears that the order-by-order improvement in the predictions of the differential cross section and $P_b$ polarization deteriorates and \NNLO{} can perform better than \NNNLO. This effect is visible also at the level of phase shifts shown in \cref{fig:non_pert_phase_shifts}. It is not clear at the moment if this is due to overfitting and (or) an underlying issue with the MWPC that we employ. Our \NNNLO{} predictions are certainly influenced by the adopted values of sub-leading $\pi N$ LECs~\cite{Siemens:2016jwj}. Calculations of other scattering cross observables show that the order-by-order convergence demonstrated in \cref{fig:np_obs_ang} is representative for all elastic \np{} scattering observables in the PC by Long and Yang. Two-pion exchange is clearly important for achieving a realistic description of scattering observables with $T_\mathrm{lab} \lesssim 100$~MeV.

\begin{figure*}
    \centering
    \includegraphics[width=0.99\textwidth]{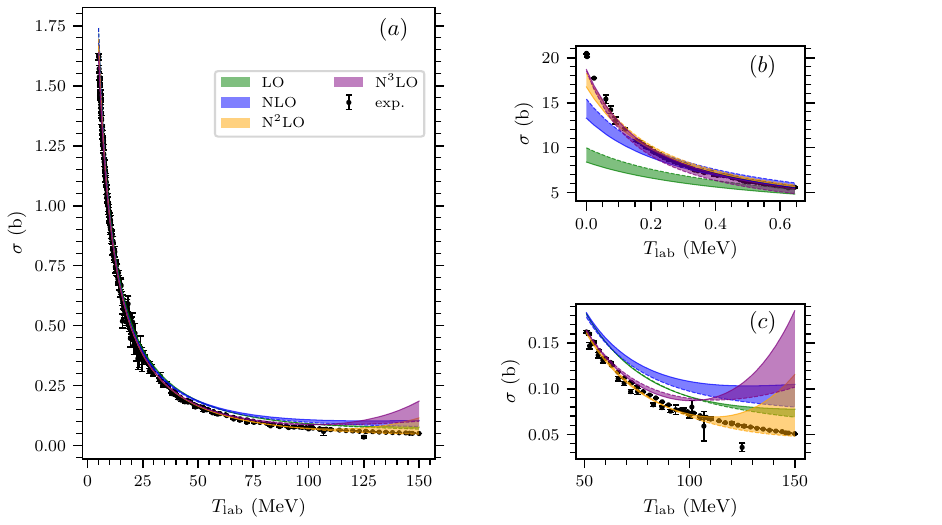}
    \caption{Total \np\ cross sections computed by integrating the differential cross sections \eqref{eq:SGT}. Panel $(a)$ shows cross sections for a large interval of scattering energies, $T_\mathrm{lab} = 5\text{--}150$~MeV. Panels $(b)$ and $(c)$ expand results at low- and high-energy intervals, respectively. The bands indicate cutoff variation as in \cref{fig:non_pert_phase_shifts}. Experimental data from Refs. \cite{Granada_1,Granada_2}.}
    \label{fig:SGT_zoom}
\end{figure*}

The total cross section can be straightforwardly computed from the differential cross section as
\begin{equation}
    \SGT(\pon) = 2\pi \int_{-1}^{1} d(\cos\thetacm) \ \frac{d\sigma}{d\Omega}(\pon,\thetacm),
    \label{eq:SGT}
\end{equation}
and predictions for scattering energies up to $T_\mathrm{lab} = 150$~MeV are shown in \cref{fig:SGT_zoom}. Also for this obvservable, the agreement with experimental data typically improves order-by-order, at least up to \NNLO. The improvement of \NNNLO{} over \NNLO{} is not obvious. At very low energies, the higher-order predictions for the total cross section are much better than the lower-order predictions. This result is somewhat peculiar for a low-energy EFT and likely due to overfitting at the phase shift level. For $T_\mathrm{lab} \gtrsim 100$~MeV, roughly corresponding to 220 MeV relative momentum, the agreement with data even deteriorates at \NNNLO. This is analogous to what was found for the angular-differential observables shown in \cref{fig:np_obs_ang} and consistent with the observation in \cref{fig:non_pert_phase_shifts} that the phase shifts at \NNNLO{} might suffer from overfitting at the higher energies. Alternatively, the observed decline in predictive power might indicate the presence of an additional mass scale at 200-300 MeV. Thus, it will be very interesting to study the effects of accounting for the $\Delta(1232)$-isobar in two-pion exchange in this MWPC.

Next, we analyze how the perturbative breaking of unitarity in $\chi$EFT affects the predictions of total cross sections. Indeed, the computation of $S$-matrix elements using \cref{eq:S-T-nu}, where the order-by-order contributions of the scattering amplitudes are summed directly to the $S$-matrix, leads to a perturbative breaking of unitarity. In contrast, amplitudes computed non-perturbatively, i.e., when the potential terms are summed before solving for the scattering amplitude (as is done in WPC), are unitary by construction. In this case, the probability flux in the scattering process is also conserved exactly and the optical theorem can be safely used to compute the total cross section as, e.g., 
\begin{equation}
    \SGT(\pon) = \frac{2\pi}{\pon}\Im\left[a(\thetacm = 0) + b(\thetacm=0)\right],
    \label{eq:SGTopt}
\end{equation}
where $a(\thetacm)$ and $b(\thetacm)$ are Saclay-amplitudes computed from the $M$-matrix \cite{Bystricky:1976jr}. 

\begin{figure}
    \centering
    \includegraphics[width=0.99\columnwidth]{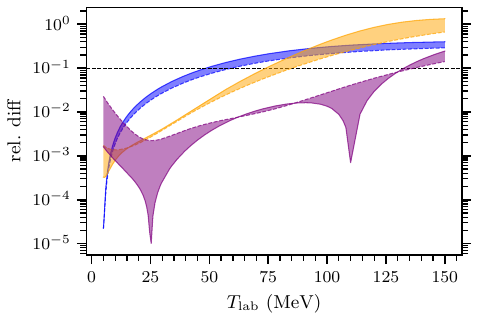}
    \caption{The relative difference between  total \np{} cross sections ($\sigma$) computed by integrating of the differential cross section~\eqref{eq:SGT} and the optical theorem~\eqref{eq:SGTopt}. The bands indicate cutoff variation as in \cref{fig:non_pert_phase_shifts}. The color coding for the orders is the same as \cref{fig:non_pert_phase_shifts}. The horizontal dashed line marks a $10$\% difference.}
    \label{fig:opt_no_opt}
\end{figure}

We use the difference between total cross sections calculated using \cref{eq:SGT} and \cref{eq:SGTopt} to measure the effects of unitarity breaking. In~\cref{fig:opt_no_opt} we show the relative difference between the cross sections computed using exact integration and the optical theorem as a function of scattering energy. The figure demonstrates how unitarity is restored perturbatively as we go to higher chiral orders. Indeed, the relative difference between the two cross section calculations is limited to 10\% for scattering energies up to 40~MeV at \NLO, 70~MeV at \NNLO, and 120~MeV at \NNNLO{}, respectively. The bands in the figure reflect differences coming from using two cutoff values 500~MeV and 2500~MeV. 
The bands for \NLO{} and \NNLO{} increase smoothly with the scattering energy. The band at \NNNLO{} shows an artifact from the two different calculations for $\Lambda = 2500$~MeV intersecting at some energies leading to very small relative errors. We also note that the cutoff dependencies for the \NNLO{} and \NNNLO{} calculations do not vanish as the scattering energy approaches zero.

We can also discuss this result in terms of the EFT truncation error. For a given chiral order, we argue that the results from the two different cross section calculations should not become significantly different until we reach an energy where the next (omitted) order in the chiral low-energy expansion becomes relevant. This should correspond to the scattering energy for which the truncation error is significant. Breaking unitarity implies that the norm of the partial-wave $S$-matrix in \cref{eq:S-T-nu} deviates from unity as $\left(S^{(\nu)}\right)^\dagger S^{(\nu)} = 1 - \mathcal{C} (Q/\Lambda_b)^{\nu+1}$, where we also expect $\mathcal{C}$ to be of natural size. This scaling of unitarity breaking should be revisited when probability distributions of the LEC values and the hyperparameters of the EFT truncation error have been inferred using a Bayesian approach.

\section{Summary and outlook\label{sec:conclusions}}
This work presents a comprehensive analysis of \np{} scattering observables (cross sections and polarizations) utilizing an RG-invariant formulation of $\chi$EFT by Long and Yang. We calibrated the LECs by reproducing Nijmegen phase shifts at specific scattering energies, and carried out calculation up to \NNNLO{} for two values of the momentum-space cutoffs, $500$ MeV and $2500$ MeV. The PC that we employed is fairly representative of a broad class of MWPCs in which corrections beyond \LO{}, based on one-pion exchange, are included perturbatively and the short-range contact potential incorporates counterterms promoted to renormalize the long-range pion contributions to the scattering amplitudes. A key result of this paper was a quantitative demonstration that RG-invariant $\chi$EFT exhibits a steady order-by-order convergence in the description of scattering observables, starting already at \LO. A second key result was the realistic reproduction of experimental scattering data in an energy range up to $T_\mathrm{lab} =100$ MeV at \NNLO. We also found that \NNNLO{} predictions do not always improve over \NNLO{}. 

A perturbative approach exposes the deficiencies of any PC, not only the possible lack of RG-independence. In fact, using a perturbative approach we found that the accuracy of our \NNNLO{} predictions for the total \np{} cross section declines as one approaches $T_\mathrm{lab} \gtrsim 100$ MeV. This corresponds to a relative scattering momentum of 220 MeV and might suggest the presence of an additional mass scale at 200--300 MeV. This finding is in accordance with the known mass splitting between the nucleon and the $\Delta$(1232) resonance, but is markedly lower than conventional estimates of the breakdown scale of $\chi$EFT residing in the vicinity of the $\rho$-meson mass. The latter estimate has also been corroborated in a Bayesian  study of non-perturbative WPC predictions of nucleon-nucleon scattering observables \cite{Melendez:2017phj}. 

Based on our comparison of perturbative and non-perturbative calculations of phase shifts, we speculated that the magnitudes of the imaginary component of the non-perturbative phase shift and the $\chi$EFT truncation error are linked. We also investigated the breaking of unitarity at the level of total $np$ cross sections. The connection between perturbative unitarity breaking and the truncation error deserves further attention.

Future work will focus on quantifying posterior probability distributions for the LECs and the EFT truncation error, making predictions beyond the two-nucleon system, and the effects of including the $\Delta$(1232) resonance in the two-pion exchange potential. Fast and accurate emulators~\cite{Duguet:2023wuh}, adapted to perturbative computations, will likely be essential for rigorous testing of RG-invariant $\chi$EFT against nuclear data and to address critical questions regarding, e.g., the construction of \LO{}, the importance of promoting higher-order pion exchanges and many-nucleon forces as one increases the mass number, and the level of fine-tuning in $\chi$EFT. 

\begin{acknowledgments}
O.T.~thanks C.-J.~Yang, B. Long, and R.~Peng for helpful discussions and for providing detailed benchmarks. The authors also thank Daniel Phillips for feedback on a draft version of the manuscript. This work was supported by the European Research Council (ERC) under the European Unions Horizon 2020 research and innovation program (Grant Agreement No.~758027), the Swedish Research Council (Grants No.~2017-04234, No.~2020-05127 and No.~2021-04507). 
\end{acknowledgments}
\FloatBarrier
\newpage
\bibliography{bib} 

\clearpage\newpage
\appendix
\begin{widetext}
\section{Nuclear potentials in the Long and Yang power counting \label{app:LY_and_pot}}
The orders at which potentials appear in the Long and Yang PC in channels where OPE is treated non-perturbatively are shown in \cref{tab:potentials_PC}. Similarly, for the channels where OPE is treated perturbatively, we follow the PC of Ref.~\cite{PhysRevC.99.024003} also shown in \cref{tab:potentials_PC}. In this appendix, we list the expressions for the potentials appearing in \cref{tab:potentials_PC}. The potential contributions will be listed using the following decomposition convention \cite{Machleidt:2011zz}
\begin{equation}
\begin{alignedat}{1}
    V(\bm{p}',\bm{p}) &= V_C + \bm{\tau}_1 \cdot \bm{\tau}_2 W_C  \\ &+ \left[ V_S + \bm{\tau}_1 \cdot \bm{\tau}_2W_S\right]\bm{\sigma}_1 \cdot \bm{\sigma}_2 + \\
    &+ \left[ V_{LS} + \bm{\tau}_1 \cdot \bm{\tau}_2 W_{LS}\right](-i\bm{S} \cdot \left(\bm{q} \times \bm{k})\right)  \\ &+ \left[ V_{T} + \bm{\tau}_1 \cdot \bm{\tau}_2 W_{T}\right]\bm{\sigma}_1\cdot \bm{q} \bm{\sigma}_2 \cdot \bm{q} \\
    &+ \left[ V_{\sigma L} + \bm{\tau}_1 \cdot \bm{\tau}_2 W_{\sigma L}\right]\bm{\sigma}_1\cdot (\bm{q}\times \bm{k}) \bm{\sigma}_2 \cdot (\bm{q}\times \bm{k}),
\end{alignedat}
\label{eq:pot_decomp}
\end{equation}
where
\begin{equation}
    \bm{q} = \bm{p}-\bm{p}', \quad \bm{k} = \frac{1}{2}\left(\bm{p}+\bm{p}'\right), \quad \bm{S} = \frac{1}{2}\left(\bm{\sigma}_1 + \bm{\sigma}_2\right)
\end{equation}
and $\bm{\sigma}_i$ denotes the Pauli spin matrix for the respective nucleon.

The one-pion exchange potential takes the form
\begin{align}
    V^{(0)}_{1\pi} &= \left(\bm{\tau}_1 \cdot \bm{\tau}_2 \right) \left(\bm{\sigma}_1\cdot \bm{q} \bm{\sigma}_2 \cdot \bm{q} \right) W_T, \\
    W_T &= -\left(\frac{g_A}{2f_\pi}\right)^2 \frac{1}{q^2 + m^2_\pi},
\end{align}
where $g_A = 1.29$ is the axial coupling, $f_\pi = 92.1$~MeV the pion decay constant, $m_\pi = 138.039$~MeV is the average pion mass and $q=|\bm{q}|$. For the two-pion exchange potentials, we employ expressions computed with dimensional regularization (DR). The leading two-pion exchange potential takes the form \cite{Epelbaum:1998ka,Epelbaum:1999dj,Machleidt:2011zz}
\begin{align}
    V^{(2)}_{2\pi} &= \bm{\tau}_1 \cdot \bm{\tau}_2 W_C +  \bm{\sigma}_1 \cdot \bm{\sigma}_2  V_S +  \bm{\sigma}_1\cdot \bm{q} \bm{\sigma}_2 \cdot \bm{q} V_T,\\
    W_C &= -\frac{L(q)}{384\pi^2 f^4_\pi}\Bigg[4m^2_\pi\left(5g^4_A-4g^2_A-1\right) + 
    q^2\left(23g^4_A-10g^2_A -1\right) + \frac{48g^4_A m^4_\pi}{w^2}\Bigg], \\
    V_S &= \frac{3g^4_A L(q) q^2}{64\pi^2 f^4_\pi}, \\
    V_T &= -\frac{1}{q^2} V_S = -\frac{3g^4_A L(q)}{64\pi^2 f^4_\pi},
\end{align}
with
\begin{equation}
    L(q) = \frac{w}{q}\ln \frac{w+q}{2 m_\pi}, \quad w = \sqrt{4m^2_\pi + q^2}.
\end{equation}
The sub-leading two-pion exchange potential takes the form of Eqs. (4.13) - (4.20) in \cite{Machleidt:2011zz}. We apply the power counting $\left(Q/\MN\right) = \left(Q/\Lambda_b\right)^2$ for $(1/m_N)$ corrections, which means that all terms proportional to $1/\MN$ vanish at order $\left(Q/\Lambda_b\right)^3$ (\NNNLO). The non-zero contributions read
\begin{align}
    V^{(3)}_{2\pi} &= V_C + \left(\bm{\tau}_1 \cdot \bm{\tau}_2 \right) \left(\bm{\sigma}_1\cdot \bm{q} \bm{\sigma}_2 \cdot \bm{q} \right) W_T, \\
    V_C &= -\frac{3g^2_A}{16\pi f^4_\pi}\Big[ 2m^2_\pi(2c_1-c_3) - q^2c_3\Big]\tilde{w}^2A(q), \\
    W_T &= -\frac{1}{q^2} W_S = -\frac{g^2_A A(q)}{32\pi f^4_\pi}c_4 w^2, \\
\end{align}
with
\begin{equation}
    A(q) = \frac{1}{2q}\arctan\frac{q}{2m_\pi},\quad \tilde{w} = \sqrt{2m^2_\pi + q^2}.
\end{equation}
For the $\pi N$ LECs $c_1,c_3,c_4$, appearing in $V^{(3)}_{2\pi}$, we employ numerical values determined in a Roy-Steiner analysis at NLO: $c_1 = -0.74$~GeV$^{-1}$, $c_3 = -3.61$~GeV$^{-1}$ and $c_4 = 2.44$~GeV$^{-1}$ \cite{Siemens:2016jwj}.

The potential contributions at each order in the channels where OPE is treated non-perturbatively are listed in \cref{tab:LYvK_LECs_up_to_N3LO}. We denote counterterms in coupled channels by a $2\times 2$ matrix representing $\l' = j\mp 1$ (rows) and $\l=j\mp 1$ (columns). \Cref{tab:LYvK_LECs_up_to_N3LO} expands upon Table I in Ref. \cite{Long:2012ve} to also explicitly show the perturbative corrections to LECs present at each order. \Cref{tab:LYvK_N_LECs} summarizes the number of LECs present at each order, excluding the three $\pi N$ LECs at \NNNLO{} from the total number.

\begin{table*}[ht]
\caption{Potential contributions at each chiral order in the channels where OPE is treated non-perturbatively. This table complements the information in \cref{tab:potentials_PC}.}
\centering

\begingroup
\renewcommand{\arraystretch}{1.5}
\begin{tabular}{c|c|c}
    Order & Pion contribution &  Contact terms \\
    \hline\hline
    \rule{0pt}{4ex}    
    \LO & $V^{(0)}_{1\pi}$ &  \multicolumn{1}{l}{$V^{(0)}_\mathrm{ct}:$}\\ 
    
    &  & $C^{(0)}_{^1S_0}$, $\begin{pmatrix} C^{(0)}_{^3S_1}& 0 \\ 0 & 0 \end{pmatrix}$, $D^{(0)}_{^3P_0}p'p$, $\begin{pmatrix} D^{(0)}_{^3P_2}p'p & 0 \\ 0 & 0 \end{pmatrix}$\\[5mm]
    \hline\hline
    \rule{0pt}{4ex}  
    \NLO & - & \multicolumn{1}{l}{$V^{(1)}_\mathrm{ct}$:} \\
    &  &   $D^{(0)}_{^1S_0}(p'^2+p^2)$, $C^{(1)}_{^1S_0}$ \\[1mm]
    \hline\hline 
    \rule{0pt}{4ex} 
   \NNLO & $V_{2\pi}^{(2)}$ &  \multicolumn{1}{l}{$V^{(2)}_\mathrm{ct}$:}\\[1mm]
    && $E^{(0)}_{^1S_0}p'^2 p^2$, $D^{(1)}_{^1S_0}(p'^2+p^2)$, $C^{(2)}_{^1S_0}$, \\[2mm]
    &  &  $\begin{pmatrix} D^{(0)}_{^3S_1}(p'^2 + p^2) &D^{(0)}_{SD} p^2\\D^{(0)}_{SD}p'^2 & 0 \end{pmatrix}$, $\begin{pmatrix} C^{(1)}_{^3S_1}& 0 \\ 0 & 0 \end{pmatrix}$,\\[5mm]
    && $E^{(0)}_{^3P_0}p'p(p'^2+p^2)$ , $D^{(1)}_{^3P_0}p'p$,  \\[2mm]
    &&  $p'p\begin{pmatrix} E^{(0)}_{^3P_2}(p'^2 + p^2) &E^{(0)}_{PF} p^2\\E^{(0)}_{PF}p'^2 & 0 \end{pmatrix}$, $\begin{pmatrix} D^{(1)}_{^3P_2}p'p & 0 \\ 0 & 0 \end{pmatrix}$, \\[5mm]
    && $D^{(0)}_{^1P_1}p'p$, $D^{(0)}_{^3P_1}p'p$ \\[1mm]
    \hline\hline
    \rule{0pt}{4ex}    
    \NNNLO & $V_{2\pi}^{(3)}$, (includes & \multicolumn{1}{l}{$V^{(3)}_\mathrm{ct}$:}\\[1mm]
    &  $\pi N$ LECs: $c_1,c_3,c_4$) & $F^{(0)}_{^1S_0}p'^2p^2(p'^2+p^2)$, $E^{(1)}_{^1S_0}p'^2 p^2$, $D^{(2)}_{^1S_0}(p'^2+p^2)$, $C^{(3)}_{^1S_0}$, \\[2mm]
    & &   $\begin{pmatrix} D^{(1)}_{^3S_1}(p'^2 + p^2) &D^{(1)}_{SD} p^2\\D^{(1)}_{SD}p'^2 & 0 \end{pmatrix}$, $\begin{pmatrix} C^{(2)}_{^3S_1}& 0 \\ 0 & 0 \end{pmatrix}$,\\[5mm]
    &&  $E^{(1)}_{^3P_0}p'p(p'^2+p^2)$, $D^{(2)}_{^3P_0}p'p$,\\
      \rule{0pt}{6ex}&&$p'p \begin{pmatrix} E^{(1)}_{^3P_2}(p'^2 + p^2) &E^{(1)}_{PF} p^2\\E^{(1)}_{PF}p'^2 & 0 \end{pmatrix}$, $\begin{pmatrix} D^{(2)}_{^3P_2}p'p & 0 \\ 0 & 0 \end{pmatrix}$, \\[5mm]
      && $D^{(1)}_{^1P_1}p'p$, $D^{(1)}_{^3P_1}p'p$ \\[2mm]
    \hline\hline
\end{tabular}
\endgroup
\label{tab:LYvK_LECs_up_to_N3LO}
\end{table*}

\begin{table}[ht]
    \caption{The number of LECs at each order in the Long and Yang PC.}
\centering
\begin{tabular}{c|c|c|c}
    Chiral order & New LECs & Pert. correction & Total up to order \\
    \hline\hline
    \LO   & 4               & --  & 4  \\
    \NLO  & 1               & 1  & 6  \\
    \NNLO & 8               & 5  &19 \\
    \NNNLO & 1 (+3\footnote{Sub-leading $\pi N$ LECs: $c_1,c_3,c_4$ excluded from the total in the last column.})  & 13 & 33 \\
    \hline
\end{tabular}
\label{tab:LYvK_N_LECs}
\end{table}

\section{Numerical implementation of distorted-wave perturbation theory\label{app:numerical}}
This appendix gives some more details regarding the implementation of the equations for higher-order corrections to the scattering amplitude in \cref{eq:TNLO,eq:TN2LO,eq:TN3LO}. Since all operator products reduce to the form in \cref{eq:fund_G0_prod}, the implementation can be done in complete analogy with the solution of the partial-wave Lippmann-Schwinger equation using Gauss-Legendre quadrature \cite{Haftel:1970zz,Landau:1990qp}.

In this appendix we suppress the conserved quantum numbers $s$ and $j$, and write the resolution of identity in the partial wave basis as
\begin{equation}
    \mathds{1} = \sum_{\l}\int_0^\infty dk \ k^2 \ket{k,\l}\bra{k,\l}.
    \label{eq:one}
\end{equation}
Furthermore, for a stationary proton (mass $m_p$) and an incoming neutron (mass $m_n$) with kinetic energy $\Tl$ in the laboratory  frame of reference, the modulus of the \CM momentum, $\pon$, is given by
\begin{equation}
    \pon^2 = \frac{m_p^2 T_\mathrm{lab} (2m_n + T_\mathrm{lab})}{(m_n+m_p)^2 + 2m_p T_\mathrm{lab}}.
    \label{eq:q_on_shell}
\end{equation}
By inserting the resolution of identity in \cref{eq:fund_G0_prod} and discretizing the integral using Gauss-Legendre quadrature with momentum points and weights, $\{k_i,w_i\}_{i=1}^N$, we obtain 
\begin{align}
    \braket{p',\l'| A_1G^+_0A_2 |p,\l} &= \sum_{\l'',\l'''} \int_0^\infty dk_1 \ k_1^2 \int_0^\infty dk_2 \ k^2_2\braket{p',\l'|A_1|k_1,\l''} \braket{k_1,\l'' |G^+_0|k_2,\l'''} \braket{k_2,\l'''| A_2 |p,\l} = \nonumber \\
    &= \sum_{\l''} \int_0^\infty dk_1 \ k_1^2 \braket{p,\l'|A_1|k_1,\l''} \frac{m_N}{\pon^2-k_1^2 + i\epsilon} \braket{k_1,\l''| A_2 |p, \l}= \\
    &= \sum_{\l''} \sum_{i=1}^N k^2_i w_i \braket{p,\l'|A_1|k_i,\l''} \frac{m_N}{\pon^2-k_i^2 + i\epsilon} \braket{k_i,\l''| A_2 |p, \l}.
\label{eq:AG0B_pw}
\end{align} 
Here, $\pon$ denotes the on-shell momentum for a given scattering energy $\Tl$ given by \cref{eq:q_on_shell}. Doing some manipulations and converting the $+i\epsilon$ prescription to a principal value we obtain \cite{Glockle,Landau:1990qp}
\begin{align}
   \braket{p',\l'| A_1G^+_0A_2 |p,\l} &= \sum_{l''} \sum_{i=1}^N k_i^2 w_i\braket{p',\l'|A_1|k_i,\l''} \frac{m_N}{\pon^2-k_i^2} \braket{k_i,\l''| A_2 | p,\l}  \nonumber \\
    &-\braket{p',\l'|A_1|\pon,\l''}\braket{\pon,\l''| A_2 | p,\l}  
    \left[m_N \pon^2 \sum_{i=1}^N \frac{w_i}{\pon^2-k_i^2} + \frac{i\pi m_N \pon}{2} - m_N \pon \ \mathrm{arctanh}\left(\frac{\pon}{\tilde{\Lambda}}\right) \right].
    \label{eq:AG0B_full_sum}
\end{align}    
All potentials are regulated using \cref{eq:inc_reg} and at sufficiently high momentum, $\tilde{\Lambda}$, all potential matrix elements are essentially zero. This means that the integral in \cref{eq:AG0B_pw} is well represented by the discretized sum where the momentum points and weights $\{k_i,w_i\}_{i=1}^N$ are chosen using Gauss-Legendre quadrature in the interval $[0,\tilde{\Lambda}]$. The last term in the bracket in \cref{eq:AG0B_full_sum} implements the principal-value integral on the interval $[\tilde{\Lambda},\infty]$ analytically since the grid is just doing the integration on $[0,\tilde{\Lambda}]$ \cite{Hoppe:2017lok}. It is possible to have a grid that extends to numerical infinity, but this generally leads to slower convergence with $N$. For the calculations in this study, we employ $\tilde{\Lambda} = \Lambda + 1500$~MeV, for both $\Lambda = 500$~MeV and $\Lambda = 2500$~MeV, which we find sufficient for numerical convergence.

\Cref{eq:AG0B_full_sum} can be expressed in a simpler form using matrix products, which speeds up the computations. We define the propagator matrix as
\begin{equation}
    [G^+_0]_{ij} = \delta_{ij} F_i, \quad     F_i = \begin{cases} \frac{m_N}{\pon^2-k_i^2}, \quad i = 1,...,N \\
    -f(\pon), \quad i=N+1, \end{cases}
\end{equation}
where
  \begin{equation}
    f(\pon) = m_N \pon^2 \sum_{i=1}^N \frac{w_i}{\pon^2-k_i^2} + \frac{i\pi m_N \pon}{2} - m_N \pon \ \mathrm{arctanh}\left(\frac{\pon}{\tilde{\Lambda}}\right).
\end{equation}  
Similarly, we make the following definitions of matrices for $A_\mu$, $\mu=1,2$,
 \begin{alignat}{2}
    &[A_\mu^{\l'\l}]_{i, j}         &&= k_i \sqrt{w_i}\braket{k_i,\l'|A_\mu|k_j,\l} k_j \sqrt{w_j}, \quad i,j=1,\dots,N \\
    &[A_\mu^{\l'\l}]_{i, j=N+1}     &&= k_i \sqrt{w_i} \braket{k_i,\l'|A_\mu|\pon,\l},\quad i=1,\dots,N   \\
    &[A_\mu^{\l'\l}]_{i=N+1, j}     &&= \braket{\pon,\l'|A_\mu|k_j,\l} k_j \sqrt{w_j},\quad j=1,\dots,N  \\
    &[A_\mu^{\l'\l}]_{i=N+1, j=N+1} &&= \braket{\pon,\l'|A_\mu|\pon,\l}, 
\end{alignat}   
effectively including an extra momentum-grid point $k_{N+1} \equiv \pon$ with weight $\sqrt{w_{N+1}} = \pon^{-1}$. Using these definitions and defining $D = A_1G^+_0A_2$, \cref{eq:AG0B_full_sum} can be written using $(N+1)\times (N+1)$ matrix products
  \begin{equation}
    [D^{\l'\l}]_{ij} = \sum_{\l''} \sum_{n,m = 1}^{N+1} [A_1^{\l'\l''}]_{in} [G^+_0]_{nm}[A_2^{\l''\l}]_{mj}, \quad i,j = 1,\dots,N+1.
    \label{eq:D_matrix}
\end{equation}  
For coupled channels, we further eliminate the sum over $\l''$ in \cref{eq:D_matrix} by defining $(2N+2) \times (2N+2)$ block-matrices, which for $A_1$ reads
\begin{equation}
    [\bm{A}_1] = \begin{pmatrix} [A_1^{--}] & [A_1^{-+}] \\ [A_1^{+-}] & [A_1^{++}] \end{pmatrix}.
    \label{eq:A_matrix}
\end{equation}
The $\pm$ notation represents $\l=j\pm1$. The propagator is diagonal in $\l$ and can be written as
\begin{equation}
    [\bm{G}^+_0] = \begin{pmatrix} [\bm{G}^+_0] & 0 \\ 0 & [\bm{G}^+_0] \end{pmatrix}.
\end{equation}
We can finally write \cref{eq:D_matrix} as
\begin{equation}
    [\bm{D}] = [\bm{A}_1][\bm{G}^+_0][\bm{A}_2].
    \label{eq:matrix_prod}
\end{equation}

Note that the simplification of \cref{eq:AG0B_full_sum} to an ordinary matrix product in \cref{eq:matrix_prod} is only possible due to the specific structure of having $G^+_0$ in between $A_1$ and $A_2$. This structure gives rise to the last ``on-shell'' term in \eqref{eq:AG0B_full_sum} that can be incorporated by adding the grid point $k_{N+1} = \pon$,  which then extends the sum in \cref{eq:AG0B_full_sum} to $N+1$. \Cref{eq:D_matrix} can now be used recursively to compute longer products such as $\braket{p',\l'|A_1G^+_0A_2G^+_0A_3|p, \l}$.

As an example, the first-order correction to the $T$-matrix in \cref{eq:TNLO} can be expressed as the matrix equation
\begin{equation}
    [\bm{T}^{(1)}] =\left(\mathds{1} + [\bm{T}^{(0)}][\bm{G}^+_0]\right)[\bm{V}^{(1)}] \left(\mathds{1} + [\bm{G}^+_0][\bm{T}^{(0)}]\right).
\end{equation}

\section{Perturbative phase shifts\label{app:pert_phase}}
In this appendix we discuss how to obtain phase shifts given perturbative corrections to the $T$-matrix computed from \cref{eq:TNLO,eq:TN2LO,eq:TN3LO}. We will follow the method outlined in Ref.~\cite{PhysRevC.85.034002} and add some additional details. For uncoupled scattering channels, the $1\times 1$ $S$-matrix can be parameterized by
\begin{equation}
    S = \exp\left(2i\delta\right),
\end{equation}
where $\delta$ is the phase shift. We expand both the phase shifts and the on-shell $S$-matrix with the contributions at each chiral order obtaining
\begin{align}
    &S^{(0)} +  S^{(1)} +  S^{(2)} +  S^{(3)} +\mathcal{O}(Q^3) = \\
    & \exp\left(2i\left[\delta^{(0)} +  \delta^{(1)} +  \delta^{(2)} +  \delta^{(3)} + \mathcal{O}(Q^3)\right]\right).
\end{align}   
Performing a Taylor expansion of both sides, and matching chiral orders, gives 
\begin{align}
    S^{(0)} &= \exp\left(2i\delta^{(0)}\right) \\
    S^{(1)} &= 2i\delta^{(1)} \exp\left(2i\delta^{(0)}\right)\\
    S^{(2)} &= \left[2i\delta^{(2)} - 2 \left(\delta^{(1)}\right)^2 \right]\exp\left(2i\delta^{(0)}\right)\\
    S^{(3)} &= \left[2i\delta^{(3)} - 4 \delta^{(1)}\delta^{(2)} - \frac{4i}{3}\left(\delta^{(1)}\right)^3\right]\exp\left(2i\delta^{(0)}\right)
\end{align}
From these equations, we straightforwardly obtain explicit expressions for the \LO{} phase shift $\delta^{(0)}$ (trivial), and all corrections $\{\delta^{(\nu)}\}_{\nu>0}$. We note that all corrections are real valued. To obtain the total phase shift at, e.g., \NNLO, one has to sum $\delta^{(0)} + \delta^{(1)}+\delta^{(2)}$. The $S$-matrix corrections are obtained from the $T$-matrix corrections as
\begin{equation}
    S^{(\nu)}_{\l'\l} = -i\pi m_N \pon T^{(\nu)}_{\l'\l}, \quad \nu>0,
\end{equation}
for a given on-shell momentum, $\pon$.

For coupled channels we use the Stapp-parametrization \cite{Stapp:1956mz} for the on-shell $2\times 2$ $S$-matrix 
\begin{equation}
    S = \begin{pmatrix} \cos(2\epsilon) e^{2i \delta_1} & i \sin(2\epsilon)e^{i(\delta_1 + \delta_2)} \\ i \sin(2\epsilon)e^{i(\delta_1 + \delta_2)} & \cos(2\epsilon) e^{2i \delta_2} \end{pmatrix},
    \label{eq:S_Stapp}
\end{equation}
where the three phase shifts $\delta_1$, $\delta_2$ and $\epsilon$ parameterize the amplitude for a given channel. We now proceed completely analogous to the uncoupled case, dividing the $S$-matrix and phase shifts into chiral orders as
 \begin{equation}
    S = \sum_{\nu=0}^\infty = S^{(\nu)}, \quad
    \delta_1 = \sum_{\nu=0}^\infty \delta^{(\nu)}_1, \quad
    \delta_2 = \sum_{\nu=0}^\infty \delta^{(\nu)}_2, \quad
    \epsilon = \sum_{\nu=0}^\infty \epsilon^{(\nu)}. 
    \label{eq:order_exp}
\end{equation}   
For convenience, we define the functions
\begin{align}
    f_{11}(\epsilon,\delta_1) &= \cos(2\epsilon) e^{2i\delta_1}, \\
    f_{12}(\epsilon, \delta_1,\delta_2) &= i\sin(2\epsilon) e^{i(\delta_1 + \delta_2)}, \\
    f_{22}(\epsilon,\delta_2) &= \cos(2\epsilon) e^{2i\delta_2},
\end{align}
which are the constituents of the matrix in \cref{eq:S_Stapp}. Inserting the expansions in \cref{eq:order_exp} into \cref{eq:S_Stapp}, Taylor expanding and matching chiral orders, gives the perturbative corrections to the phase shifts. Expanding the upper left matrix element of $S$ gives
\begin{align}
    S_{11}^{(0)} &= f_{11} \\
    S_{11}^{(1)} &= \partial_\epsilon f_{11} \times \epsilon^{(1)} + \partial_\delta f_{11} \times \delta^{(1)} \\ 
    S_{11}^{(2)} &= \partial_\epsilon f_{11} \times \epsilon^{(2)} + \partial_\delta f_{11} \times \delta^{(2)}  \nonumber \\
    & + g^{(2)}_{11}(\epsilon^{(1)},\delta^{(1)}) \\
    S_{11}^{(3)} &= \partial_\epsilon f_{11} \times \epsilon^{(3)} + \partial_\delta f_{11} \times \delta^{(3)} \nonumber \\
    &+ g^{(3)}_{11} (\epsilon^{(1)},\delta^{(1)},\epsilon^{(2)},\delta^{(2)})
\end{align}
where the functions $g^{(\nu)}_{11}$ are introduced to capture all non-linear terms in the expansion
\begin{align}
    g^{(2)}_{11}(\epsilon^{(1)},\delta^{(1)}) &= \frac{1}{2}\partial^2_\epsilon f_{11} \times \left(\epsilon^{(1)}\right)^2 + \frac{1}{2}\partial^2_\delta f_{11} \times \left(\delta^{(1)}\right)^2 + \partial_\epsilon \partial_\delta f_{11} \times \delta^{(1)}\epsilon^{(1)} \label{eq:g2_11}\\
    g^{(3)}_{11}(\epsilon^{(1)},\delta^{(1)},\epsilon^{(2)},\delta^{(2)}) &= \partial^2_\epsilon f_{11}\left(\epsilon^{(1)}\epsilon^{(2)}\right) +  \partial_\epsilon \partial_\delta f_{11}\left( \epsilon^{(1)}\delta^{(2)} +\epsilon^{(2)}\delta^{(1)}\right)\nonumber\\ &+ \partial^2_\delta f_{11}\left( \delta^{(1)}\delta^{(2)}\right) + \frac{1}{6} \partial^3_\epsilon f_{11} \left(\epsilon^{(1)}\right)^3 + 
    \frac{1}{2}\partial_\delta \partial^2_\epsilon f_{11} \left(\epsilon^{(1)}\right)^2 \delta^{(1)}\nonumber\\ &+\frac{1}{2}\partial^2_\delta \partial_\epsilon f_{11} \epsilon^{(1)} \left(\delta^{(1)}\right)^2 + \frac{1}{6} \partial^3_\delta f_{11} \left(\delta^{(1)}\right)^3.
    \label{eq:g3_11}
\end{align}    
Since $f_{11}$ depends on $\epsilon$ and $\delta_1$ the index one is suppressed. The function $f_{11}$ and all its derivatives are evaluated at $(\epsilon^{(0)},\delta^{(0)}_1)$.

For the lower right matrix element described by $f_{22}$ the expressions are completely analogous to \cref{eq:g2_11,eq:g3_11}, but with $\delta_2$ instead of $\delta_1$. For the off-diagonal elements we get
\begin{align}
    S_{12}^{(0)} &= f_{12} \\
    S_{12}^{(1)} &= \partial_\epsilon f_{12} \times \epsilon^{(1)} + \partial_{\delta_1} f_{12} \times \delta^{(1)}_1 + \partial_{\delta_2} f_{12} \times \delta^{(1)}_2 \\ 
    S_{12}^{(2)} &= \partial_\epsilon f_{12} \times \epsilon^{(2)} + \partial_{\delta_1} f_{12} \times \delta^{(2)}_1 + \partial_{\delta_2} f_{12} \times \delta^{(2)}_2 + g^{(2)}_{12}(\epsilon^{(1)},\delta^{(1)}_1,\delta^{(1)}_2) \\
    S_{12}^{(3)} &= \partial_\epsilon f_{12} \times \epsilon^{(3)} + \partial_{\delta_1} f_{12} \times \delta^{(3)}_1 + \partial_{\delta_2} f_{12} \times \delta^{(3)}_2  + g^{(3)}_{12}(\epsilon^{(1)},\delta^{(1)}_1,\delta^{(1)}_2,\epsilon^{(2)},\delta^{(2)}_1,\delta^{(2)}_2),
\end{align}
where the functions $g^{(\nu)}_{12}$ capture the non-linear terms
\begin{align}
    g^{(2)}_{12}(\epsilon^{(1)},\delta^{(1)}_1,\delta^{(1)}_2) &= \frac{1}{2}\partial^2_\epsilon f_{12} \times \left(\epsilon^{(1)}\right)^2 + \frac{1}{2}\partial^2_{\delta_1} f_{12} \times \left(\delta^{(1)}_1\right)^2 + \frac{1}{2}\partial^2_{\delta_2} f_{12} \times \left(\delta^{(1)}_2\right)^2 \nonumber \\ 
    &+\partial_{\epsilon}\partial_{\delta_1} f_{12} \epsilon^{(1)} \delta^{(1)}_1 + \partial_{\epsilon}\partial_{\delta_2} f_{12} \epsilon^{(1)} \delta^{(1)}_2 + \partial_{\delta_1}\partial_{\delta_2} f_{12} \delta^{(1)}_1 \delta^{(1)}_2 \label{eq:g2_12}\\
    g^{(3)}_{12}(\epsilon^{(1)},\delta^{(1)}_1,\delta^{(1)}_2,\epsilon^{(2)},\delta^{(2)}_1,\delta^{(2)}_2) &= \partial^2_\epsilon f_{12} \epsilon^{(1)}\epsilon^{(2)} +\partial^2_{\delta_1} f_{12} \delta^{(1)}_1\delta^{(2)}_1+\partial^2_{\delta_2} f_{12} \delta^{(1)}_2\delta^{(2)}_2 \nonumber \\
    &+\partial_\epsilon \partial_{\delta_1} f_{12} \left(\epsilon^{(1)} \delta^{(2)}_1 +\epsilon^{(2)} \delta^{(1)}_1\right) + \partial_\epsilon \partial_{\delta_2} f_{12}\left(\epsilon^{(1)} \delta^{(2)}_2 +\epsilon^{(2)} \delta^{(1)}_2\right) \nonumber \\
    &+\partial_{\delta_1} \partial_{\delta_2} f_{12} \left(\delta^{(1)}_1 \delta^{(2)}_2 +\delta^{(2)}_1 \delta^{(1)}_2\right) \nonumber \\
    &+ \frac{1}{2}\partial^2_\epsilon \partial_{\delta_1} f_{12} \left(\epsilon^{(1)}\right)^2\delta^{(1)}_1 +
    \frac{1}{2}\partial^2_\epsilon \partial_{\delta_2} f_{12} \left(\epsilon^{(1)}\right)^2\delta^{(1)}_2\nonumber \\
    &+\frac{1}{2}\partial_\epsilon \partial^2_{\delta_1} f_{12} \epsilon^{(1)} \left(\delta^{(1)}_1\right)^2 + \frac{1}{2}\partial_{\delta_2} \partial^2_{\delta_1} f_{12} \delta^{(1)}_2 \left(\delta^{(1)}_1\right)^2 \nonumber \\
    &+\frac{1}{2}\partial_\epsilon \partial^2_{\delta_2} f_{12} \epsilon^{(1)} \left(\delta^{(1)}_2\right)^2 + \frac{1}{2}\partial_{\delta_1} \partial^2_{\delta_2} f_{12} \delta^{(1)}_1 \left(\delta^{(1)}_2\right)^2 +\nonumber  \\
    &+\partial_\epsilon \partial_{\delta_1} \partial_{\delta_2}f_{12} \epsilon^{(1)} \delta^{(1)}_1 \delta^{(1)}_2\nonumber \\
    &+ \frac{1}{6}\partial^3_\epsilon f_{12} \left(\epsilon^{(1)}\right)^3 + \frac{1}{6} \partial^3_{\delta_1} f_{12} \left(\delta^{(1)}_1\right)^3 + \frac{1}{
    6}\partial^3_{\delta_2} f_{12} \left(\delta^{(1)}_2\right)^3.
    \label{eq:g3_12}
\end{align}    
The function $f_{12}$ and all its derivatives are evaluated at $(\epsilon^{(0)},\delta^{(0)}_1,\delta^{(0)}_2)$.
Note that all the functions $g^{(\nu)}_{**}$ vanish if the \NLO{} corrections $(\delta^{(1)}_1, \delta^{(1)}_2, \epsilon^{(1)})$ are zero. This is the case for all coupled channels where OPE is treated non-perturbatively as seen in \cref{tab:LYvK_LECs_up_to_N3LO}. Furthermore, in all channels where OPE is treated perturbatively the \LO{} phase shifts are all zero, which makes many of the terms in the expressions for $g^{(\nu)}_{**}$ vanish due to vanishing derivatives. Thus, in both the perturbative and non-perturbative cases, \cref{eq:g2_11,eq:g3_11,eq:g2_12,eq:g3_12} can be simplified substantially. The phase shift corrections $(\epsilon^{(\nu)},\delta^{(\nu)}_1, \delta^{(\nu)}_2)$ for $\nu=1,2,3.$ are finally obtained by solving a system of linear equations
\begin{alignat}{2}
    \mathrm{\textbf{NLO}}:& \quad \begin{pmatrix} S_{11}^{(1)} \\ S_{12}^{(1)} \\ S_{22}^{(1)}  \end{pmatrix} &= \begin{pmatrix} \partial_\epsilon f_{11} & \partial_{\delta_1} f_{11} & 0 \\
    \partial_\epsilon f_{12} & \partial_{\delta_1} f_{12} &\partial_{\delta_2} f_{12}\\ \partial_\epsilon f_{22} & 0 &\partial_{\delta_2} f_{22} \end{pmatrix} \begin{pmatrix} \epsilon^{(1)} \\ \delta^{(1)}_1 \\ \delta^{(1)}_2 \end{pmatrix} \\
    \mathrm{\textbf{N$^2$LO}}:& \quad \begin{pmatrix} S_{11}^{(2)}-g^{(2)}_{11} \\ S_{12}^{(2)} - g^{(2)}_{12}\\ S_{22}^{(2)}-g^{(2)}_{22}  \end{pmatrix} &= \begin{pmatrix} \partial_\epsilon f_{11} & \partial_{\delta_1} f_{11} & 0 \\
    \partial_\epsilon f_{12} & \partial_{\delta_1} f_{12} &\partial_{\delta_2} f_{12}\\ \partial_\epsilon f_{22} & 0 &\partial_{\delta_2} f_{22} \end{pmatrix} \begin{pmatrix} \epsilon^{(2)} \\ \delta^{(2)}_1 \\ \delta^{(2)}_2 \end{pmatrix} \\
    \mathrm{\textbf{N$^3$LO}}:& \quad \begin{pmatrix} S_{11}^{(3)}-g^{(3)}_{11} \\ S_{12}^{(3)} - g^{(3)}_{12}\\ S_{22}^{(2)}-g^{(3)}_{22}  \end{pmatrix} &= \begin{pmatrix} \partial_\epsilon f_{11} & \partial_{\delta_1} f_{11} & 0 \\
     \partial_\epsilon f_{12} & \partial_{\delta_1} f_{12} &\partial_{\delta_2} f_{12}\\ \partial_\epsilon f_{22} & 0 &\partial_{\delta_2} f_{22} \end{pmatrix} \begin{pmatrix} \epsilon^{(3)} \\ \delta^{(3)}_1 \\ \delta^{(3)}_2 \end{pmatrix}.
\end{alignat}
\end{widetext}

\end{document}